\documentclass[manuscript]{emulateapj}
\usepackage{natbib}
\usepackage{graphicx}
\usepackage{txfonts}

\shorttitle{RXTE observations and state transition}
\shortauthors{Radhika et al.}

\begin{document}

\title{RXTE observations and state transition in MAXI J1836$-$194}

\author{\textbf{Radhika. D}} 
\affil{Space Astronomy Group, ISRO Satellite Centre, Old Airport Road, Bangalore, INDIA}
\affil{Department of Physics, University of Calicut, Kerala, INDIA}
\email{radhikad\_isac@yahoo.in}

\author{\textbf{M. C. Ramadevi}}
\affil{Space Astronomy Group, ISRO Satellite Centre, Old Airport Road, Bangalore, INDIA}

\and
\author{\textbf{S. Seetha}}
\affil{Space Science Office, ISRO Headquarters, Bangalore, INDIA}

\begin{abstract}

We present the results of analysis of the X-ray transient source MAXI J1836$-$194 during its outburst in August 2011. MAXI GSC detected the source on 30th August 2011, when it started rising from the quiescence. We have studied the source using the observations of RXTE. In this paper, we study the temporal and spectral evolution of the source during the outburst. Spectral analysis shows that the source exhibits state transition to Hard Intermediate state(HIMS) and decays back to a Low/Hard state. The temporal analysis indicates the presence of QPOs during the hard intermediate state. We also observe correlation between the evolution of break frequency with respect to the spectral characteristics. We conclude that this is probably the second source after H 1743$-$322 which exhibits transition to HIMS but does not reach soft spectral state. 

\end{abstract}


\section{Introduction}

Most of the Black hole X-ray binaries are transient in nature. They are observed to remain inactive for a long time and occasionally show a sudden increase in their luminosities, which are referred to as outbursts. They achieve a peak flux and then decay back to the quiescent level. During outbursts, Black hole transients also enter different spectral states and undergo state transitions \citep{RM06}. The transient source MAXI J1836$-$194 was first detected by MAXI GSC at RA(J2000) = 279.12$^{\circ}$ and Dec(J2000) = -19.41$^{\circ}$ \citep{Negoro}. The source was also detected by SWIFT BAT in the 15 - 50 keV band \citep{Negoro}. Follow-up observations were done by SWIFT and an optical counterpart of magnitude m$_v$ = 16.20 +/- 0.04 and m$_u$ = 16.36 +/- 0.04 was found from the UVOT analysis \citep{Kennea}. RXTE observations of the source began from 31st August 2011 with its PCA instrument. Different observations were carried out in the optical and radio bands that confirmed the presence and evolution of the source(\citealt{3619}, \citealt{3628}, \citealt{3640}). The first radio lightcurve was obtained using RATAN-600 during the period from 2nd September to 22nd September 2011 \citep{Trushkin}.

In this paper we study the temporal and spectral variation of the source. We attempt to understand the evolution of the Hardness-Intensity Diagram (HID) and the variation of the spectral parameters to interpret the spectral states observed. We study the evolution of Power Density Spectra (PDS) to look for the temporal characteristics.

\section{Observations and Data Analysis}
\label{sec:oda}
\begin{figure}[t!]
\includegraphics[width=9cm]{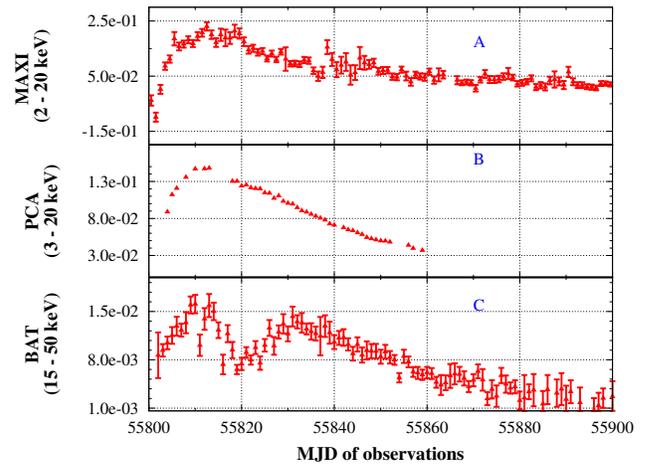}
\caption{Lightcurve of MAXI J1836$-$194 in units of photons cm$^{-2}$ sec $^{-1}$. Panel A shows the variation obtained from MAXI (2 - 20 keV) observations, panel B depicts the PCA (3 - 20 keV) flux variation and in the panel C is the lightcurve obtained from BAT (15 - 50 keV) observation.}
\label{lc}
\end{figure}

We have analysed public archival data sets obtained from the HEASARC database, of observations by RXTE in order to study the spectral and temporal characteristics of the source MAXI J1836$-$194. Figure \ref{lc} shows the lightcurve of the source obtained from the observations of MAXI, PCA and SWIFT from the beginning of outburst on MJD 55800 to MJD 55890 by which time the source has already reached back to quiescence. 

We analysed the PCA data which spans around 56 days of observations from 31st August 2011 (MJD 55804) to 25th October 2011 (MJD 55860, nearly the end of outburst). We present the lightcurve, timing and spectral analysis done for the source in the following sections. The standard FTOOLS of \textbf{HEASOFT v 6.9} and the IDL based customized software \textit{\textbf{General High energy Aperiodic Timing Software (GHATS)}} \footnote{http://www.brera.inaf.it/utenti/belloni/GHATS\_Package/Home.html} are used for the purpose.

\subsection{Timing analysis}
\label{sec:ta}

We have studied the temporal variation of the source MAXI J1836$-$194, using the Single Bit(SB) mode data of the PCA. The SB mode has a time resolution of 125 $\mu$sec. 
The Power Density Spectra (PDS) are generated with the help of GHATS, over the energy range of 2 - 20 keV, for a time resolution of 0.0078 s corresponding to a Nyquist frequency of 64 Hz. The GHATS package takes into account the dead time effect and hence the value of normalisation factor, during the subtraction of Poisson noise (as in \citealt{Zhang}) while generating the PDS. The power obtained in the PDS has units of rms$^2$/Hz. The PDS is modeled using different components of Lorentzians (as in \citealt{Belloni2002}). For the observations where QPOs are present, we obtain the values for the QPO by fitting a Lorentzian whose centroid frequency is quoted as the QPO frequency. Other characteristics of the QPO like Q - factor and significance are estimated. We estimate the rms value (amplitude) of the QPO using the {\it stat} command in QDP. A low frequency break in the PDS is observed for almost all the observations and we estimate the low break frequency by modeling the profile with a Lorentzian. We also estimate the total fractional variability (rms of PDS) over 0.01 to 64 Hz with the help of the {\it stat} command. The error values on the parameters are estimated at 90\% confidence interval (2$\sigma$), using the {\it fit err} command.

\subsection{Spectral analysis}
\label{sec:sa}
\textit{Standard 2} data products of PCU2, which provides 16 s time resolution are considered for extracting the PCA spectral data. \textit{pcabackest} tool is used for estimating the PCA background and the instrumental model files for \textbf{Epoch 5c} were obtained from the RXTE PCA background webpage\footnote{http://heasarc.gsfc.nasa.gov/docs/xte/pca\_bkg\_epoch.html}. The respective response matrices are generated for the observations using \textit{pcarsp}, applying the appropriate calibration. Spectral analysis is performed using the HEASOFT package \textbf{XSPEC v 12.6}. For a few of the observations, we also studied the HEXTE spectra which were generated following the guidelines mentioned in the RXTE news release \footnote{http://heasarc.gsfc.nasa.gov/docs/xte/whatsnew/newsarchive\_2010.html}

\begin{figure}[t!]
\includegraphics[height=8cm,width=6cm,angle=-90]{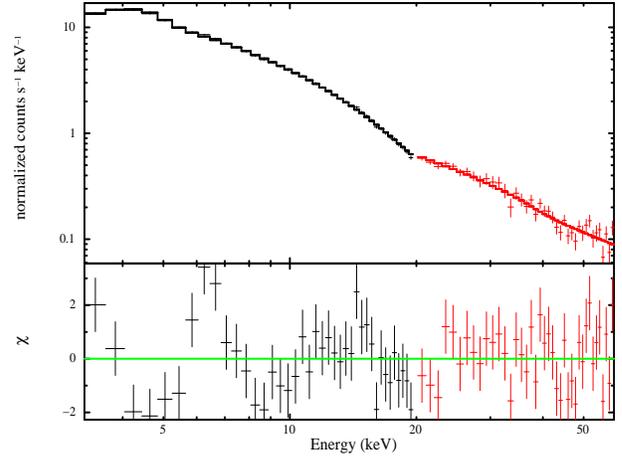}
\caption{Combined PCA and HEXTE Spectral fit for the observation ID 96371-03-01-00, on MJD 55804. Residuals can be observed at 6.4 keV.}
\label{day1-fit}
\end{figure}

A systematic error of 1\% is applied in order to take into account the uncertainities in the data. From an initial fit to the data sets we find an average value of 0.25 $\times$ 10$^{22}$ atoms cm$^{-2}$ for the n$_H$ factor to account for the interstellar absorption. This value agrees quite well with the Leiden/Argentine/Bonn (LAB) survey \citep{LAB} value of 0.22 $\times$ 10$^{22}$ atoms cm$^{-2}$ and the Dickey \& Lockman (DL) survey \citep{DL} value of 0.23 $\times$ 10$^{22}$ atoms cm$^{-2}$ \footnote{http://heasarc.gsfc.nasa.gov/cgi-bin/Tools/w3nh/w3nh.pl}. The 3 - 20 keV PCA spectrum is modeled by a multicolored disk model `diskbb' (\citealt{Mitsuda84} and \citealt{Makishima86}), along with a `powerlaw' component to consider the high energy flux contribution. The parameters of the diskbb model are, the temperature at inner disk radius T$_{in}$ and the norm factor which is related to the disk radius. For the first RXTE observation, initially we modeled the PCA spectrum, and then the combined PCA and HEXTE spectrum by \textit{phabs*(powerlaw)}, which resulted in large residuals and reduced $\chi^{2}$ of 3.71. An inclusion of thermal component using diskbb, improved the fits with reduced $\chi^{2}$ of 1.57 (see Figure \ref{day1-fit}). The residuals to the fit clearly indicates that a Gaussian component is required with line Energy $\sim$ 6.4 keV. Hence the final model used is \textit{phabs*(diskbb+gauss+powerlaw)}. PCA observations upto MJD 55849 are modeled in a similar way, while those during the later part of the outburst do not require a disk component. Thus we obtain the spectral parameters of \textit{inner disk temperature T$_{in}$, normalisation of diskbb, photon index, normalisation of powerlaw, line Energy, width of line and norm of Gaussian line}. We estimated the flux contribution by the different spectral components using \textit{cflux} command. In order to find the Hardness ratio, the flux contributions over 4 - 10 keV and 10 - 20 keV energy bands are obtained. For all the spectral parameters and flux values, errorbars are estimated at 90\% confidence interval (2$\sigma$) using \textit{err} command.

\section{Results}
\label{sec:res}
In this section, we present the results of temporal and spectral analysis of the source MAXI J1836$-$194. The lightcurve of the source MAXI J1836$-$194 from discovery to the time of quiescence, is shown in the Figure \ref{lc}. Panel A shows lightcurve obtained from \textbf{MAXI GSC} \citep{Mat2009} over the energy range of 2 - 20 keV. In panel B is the \textbf{RXTE PCA} flux variation obtained from the spectral fits to the data as in section \ref{sec:spectra} for 3 - 20 keV and in the bottom panel C is the \textbf{SWIFT BAT} lightcurve over the energy range of 15 - 50 keV.

The MAXI lightcurve (panel A) has a profile of Fast Rise and Exponential Decay (FRED) as seen in many of the Black hole transients in the soft X-ray band (\citealt{Chen}). We see that the flux rises steeply till MJD 55806 and then has a slower increase towards MJD 55812. The region between MJD 55812 and 55820, appears like a plateau phase. An exponential fit to the MAXI lightcurve gives a decay time of 41 days.

The PCA lightcurve (panel B) shows a rise of the source upto MJD 55810, followed by an exponential decay. It may be noted that the PCA had no observations during most part of the plateau phase observed in MAXI during MJD 55812 to MJD 55820. Had this plateau not been there, the whole lightcurve profile would have been a typical FRED. The PCA lightcurve gives a rise time of 9 days for the source to attain its peak flux from MJD 55804 to MJD 55813. The decay time is estimated to be of 32 days. 

The SWIFT BAT lightcurve in the energy band of 15 - 50 keV (panel C) exhibits that, the flux of hard photons initially did rise to a peak upto MJD 55810, but the flux reduced later to a minimum on MJD 55820. The hard flux is again observed to increase upto MJD 55830, although not upto the same level as previous rise on MJD 55810. Hence a double peak is seen in the lightcurve profile of the hard flux over the range of $\sim$ 30 days. The decay of the hard flux begins after MJD 55830 with a decay time of 36 days. A comparison of all three lightcurves implies that the contribution of soft photons actually peaked around MJD 55820 (panel A and C). 

\subsection{Temporal evolution}
\label{sec:te}
\begin{figure}[t!]
\includegraphics[height=8cm,width=6cm,angle=-90]{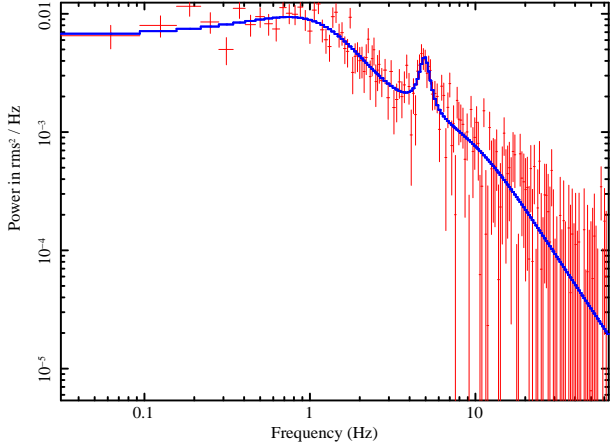}
\caption{Power spectra for one of the observations of MAXI J1836$-$194 on MJD 55819 fitted using multiple Lorentzians, shows a significant QPO}
\label{pds_15sep}
\end{figure}

\begin{figure}[t!]
\includegraphics[width=9cm]{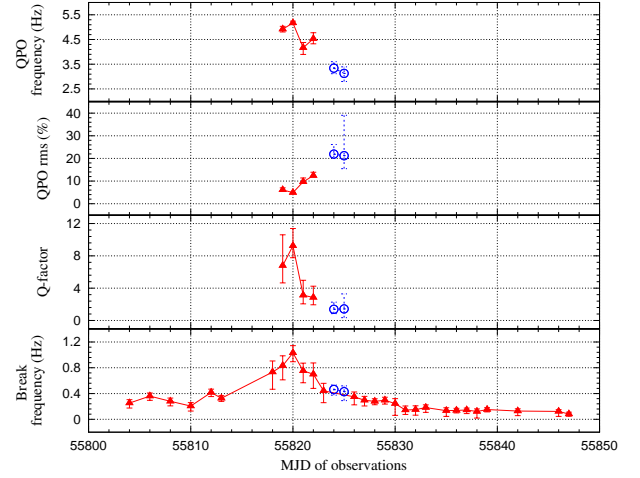}
\caption{Variation of break frequency, QPO frequency observed, the amplitude (rms) of the QPOs and their Q-factor. The triangular points in top 3 panels corresponds to MJD 55819 to 55822 and circular points are for MJD 55824 and 55825}
\label{qpo_evo}
\end{figure}

As explained in \S \ref{sec:ta}, we studied the PDS of the source MAXI J1836$-$194 over 0.01 - 64 Hz. Observations since beginning of the outburst i.e. from MJD 55804 to MJD 55818 have only a broad-band noise in the PDS, without any signature of QPO. During the observations from MJD 55819 to MJD 55822, low frequency QPOs of $\sim$ 4 to 5 Hz have been observed. In Figure \ref{pds_15sep} we show the PDS obtained for the observation on MJD 55819, over the frequency range of 0.01 to 64 Hz. For the observation on MJD 55823, no QPO is observed. Two more observations on MJD 55824 and MJD 55825 show the presence of QPOs of frequency $\sim$ 3Hz. Figure \ref{qpo_evo} shows the variation in the QPO frequency, rms, Q-factor and low break frequency. Table \ref{t1} summarizes the properties of the QPOs observed. The Q-factor is found to be $>$ 2 for observed QPOs. The different parameters of the QPOs suggest that all the QPOs observed are of Type C (see \citealt{Casella2004} for details on types of QPOs). The break frequency is observed to increase from 0.25 Hz on MJD 55804 to a maximum of 1.03 Hz on MJD 55820, and then decrease to 0.08 Hz on MJD 55847. We attempt to explain the observed temporal variations in QPO and break frequency under the discussions (section \ref{sec:discu}).

\begin{table*}
\begin{center}
\begin{tabular}{lccccc}
\hline
MJD & $\nu$ & rms & Q & Sig & Break F \\
    & (Hz)    & (\%)  &   &     & (Hz)    \\
\hline \hline \\

55819 & 4.92 \tiny{${^{+0.26}_{-0.12}}$} & 6.68\tiny{${^{+1.33}_{-0.52}}$} &  6.22 \tiny{${^{+3.9}_{-2.4}}$} & 2.44 & 0.83\tiny{${^{+0.22}_{-0.37}}$}\\\\
55820 & 5.17 \tiny{${^{+0.12}_{-0.06}}$} & 4.56\tiny{${^{+0.40}_{-0.14}}$} & 10.41 \tiny{${^{+4.4}_{-2.6}}$} & 3.10 & 1.03\tiny{${^{+0.13}_{-0.24}}$}\\\\
55821 & 4.21 \tiny{${^{+0.32}_{-0.18}}$} & 3.15\tiny{${^{+0.66}_{-0.10}}$} & 10.00 \tiny{${^{+6.3}_{-2.7}}$} & 1.56 & 0.75\tiny{${^{+0.18}_{-0.30}}$} \\\\
55822 & 4.46 \tiny{${^{+0.46}_{-0.17}}$} & 8.18\tiny{${^{+1.57}_{-0.32}}$} &  4.22 \tiny{${^{+2.6}_{-1.2}}$} & 1.40 & 0.70\tiny{${^{+0.22}_{-0.39}}$}\\\\
55824 & 3.19 \tiny{${^{+0.29}_{-0.15}}$} & 7.48\tiny{${^{+4.16}_{-1.14}}$} &  4.35 \tiny{${^{+4.5}_{-2.4}}$} & 1.61 & 0.46\tiny{${^{+0.08}_{-0.15}}$}\\\\
55825 & 3.01 \tiny{${^{+0.28}_{-0.11}}$} & 4.24\tiny{${^{+0.75}_{-0.13}}$} &  6.94 \tiny{${^{+4.2}_{-1.8}}$} & 2.16 & 0.43\tiny{${^{+0.13}_{-0.22}}$}\\\\

\hline
\end{tabular}
\\
\caption{\label{t1} Table showing the variation in temporal parameters. QPOs are observed from MJD 55819 to 55822 (triangular points in top 3 panels of Figure \ref{qpo_evo}). For two more observations on MJD 55824 and 55825 (circular points in Figure \ref{qpo_evo}), QPOs are again observed.\\ Here, $\nu$ : QPO frequency in Hz, rms : QPO amplitude in rms, Q : Q-factor of the QPO, Sig : Significance of QPO and Break F : Break Frequency of the PDS in Hz}
\end{center}
\end{table*}

\subsection{Spectral evolution}
\label{sec:spectra}

\begin{figure}[t!]
\includegraphics[width=9cm]{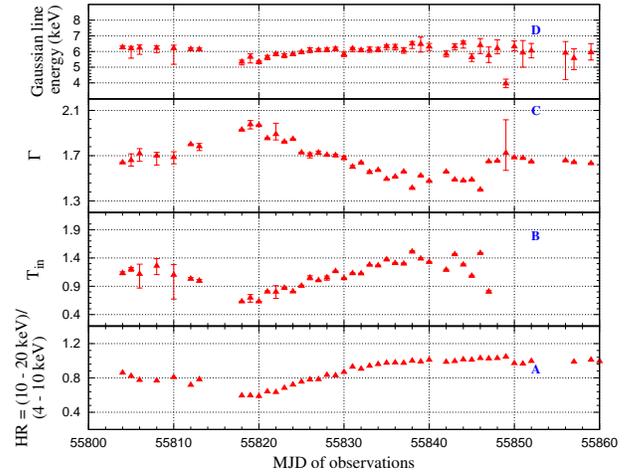}
\caption{Variation in the Hardness ratio of the source over the energy range of 10 - 20 keV w.r.t 4 - 10 keV, spectral parameters T$_{in}$, photon index $\Gamma$, energy of Fe line emission obtained from fits for RXTE PCA data in 3 - 20 keV.}
\label{spec_par}
\end{figure}

\begin{figure}[t!]
\includegraphics[width=9cm]{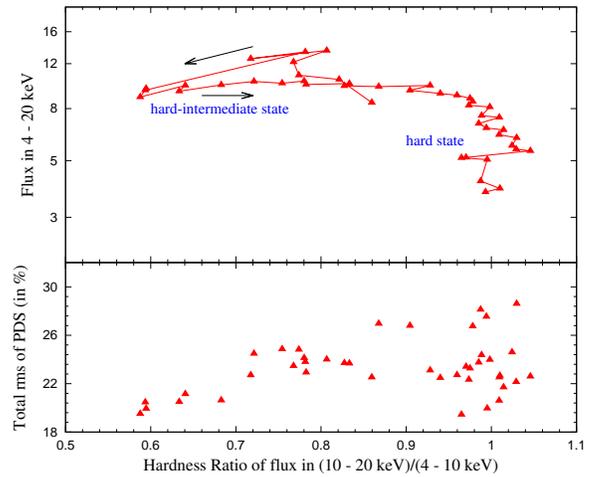}
\caption{Top panel shows the Hardness-intensity diagram (HID) and the bottom panel shows the variation of total rms of the PDS, during the outburst of the source}
\label{HID-rms}
\end{figure}

In Figure \ref{spec_par} we show the variation of the different spectral parameters like disk temperature, photon index, Fe line energy and the hardness ratio. The evolution of the Hardness-intensity diagram and the rms-hardness relation is shown in Figure \ref{HID-rms}. A detailed study on understanding the different spectral states based on the total (fractional) rms of the PDS was studied by \citealt{Munoz2011} for the Black hole source GX 339$-$4. 

During the rising phase of the outburst from MJD 55804 to MJD 55820, the inner disk temperature (panel B of Figure \ref{spec_par}) varies from 1.13 keV to 0.63 keV and the photon index (panel C of Figure \ref{spec_par}) increases from a value of 1.6 on MJD 55804 to 2.0 on MJD 55820. The decrease in hard photons is observed simultaneously as shown in panel C of Figure \ref{lc}. This suggests that the spectra is softening during the rise as the outburst progresses. The trend in the value of T$_{in}$ obtained from the fits is indicative of disk flowing inward although the data considered are above 3 keV only. Also, \citealt{Mer2000} and \citealt{Dunn2011} have already pointed out that since the parameter of disk radius does vary because of change in spectral hardening factor the diskbb model parameters are found to be not a correct representative of the spectra. During MJD 55821 to MJD 55846, the disk temperature varies from 0.8 keV to 1.4 keV while the photon index reduces from 1.8 to 1.4. This reduction in photon index implies the hardening of the spectra during the decay.  

In panel A of Figure \ref{spec_par}, we show the variation of the hardness ratio over the energy range of (10 - 20) keV to (4 - 10) keV. This energy range is chosen to match the observations made with MAXI. However since RXTE  is also sensitive $>$3keV  this energy range did not make any great difference in the interpretation of the results. The flux values over these energies are obtained after the spectral fits to each of the PCA data sets using the \textit{flux} command. We observe that during the rising phase of the outburst, the hardness ratio reduces from 0.86 (MJD 55804) to 0.58 (MJD 55820), suggesting the spectral softening. The variation in hardness can also be seen in top panel of Figure \ref{HID-rms} (see also Figure 3 of \citealt{Russell2013}). During this phase the total rms of the PDS reduces from 24\% to 19\%. From MJD 55821 to MJD 55838, the hardness ratio is observed to increase from 0.63 to 0.99. We observe that this is accompanied by the variation in the total rms of the PDS around 19\% and $\sim$25\% (bottom panel of Figure \ref{HID-rms}). The values of disk temperature, photon index, hardness ratio and total rms suggests that the source would have occupied a hard-intermediate state during this phase. From MJD 55389 onwards, the hardness ratio remains around 1 and the rms of the PDS increases to a maximum of 28\%. Also the diskbb component is not necessary for the spectral fitting. These suggest the spectral hardening and that the source enters the hard state. 

\begin{figure}[t!]
\includegraphics[width=9cm]{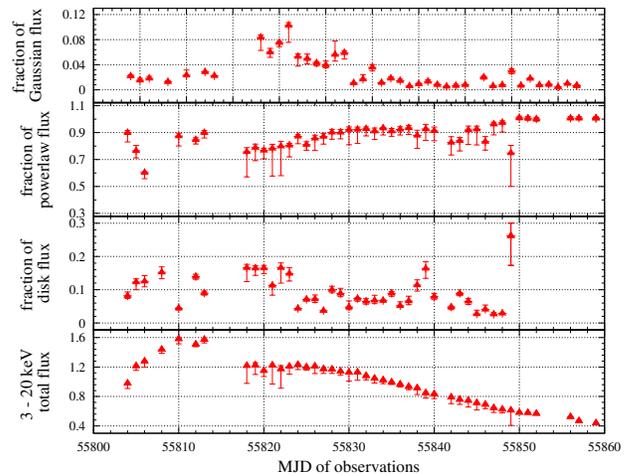}
\caption{Variation in total unabsorbed flux and fractional values of disk, powerlaw and Gaussian flux are shown.}
\label{flux-frac}
\end{figure}

\begin{figure}[t!]
\includegraphics[width=9cm]{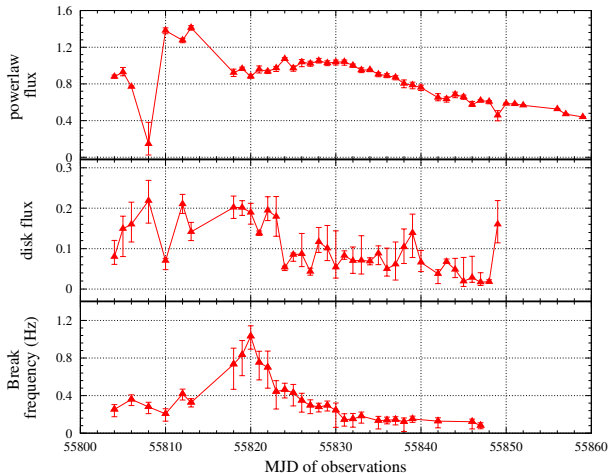}
\caption{Evolution of break frequency is shown along with variation of disk and powerlaw flux. The flux values are quoted in units of 10$^{-9}$ ergs cm$^{-2}$ sec$^{-1}$}
\label{break-flux}
\end{figure}

The variation of flux for the different model components in Figure \ref{flux-frac} shows that the powerlaw emission has a major contribution during the entire outburst with an intial contribution of $>$60\% and then dominating the entire spectra completely during the end phase of the outburst. During the initial rise from MJD 55804 to MJD 55820, both powerlaw flux and disk flux are seen to increase (see Figure \ref{break-flux}). The disk flux and the 3 - 20 keV flux becomes maximum around the peak of the outburst (top panel of Figure \ref{lc}). Thus from the spectral analysis, it is evident that the source exhibits spectral state transitions. During the initial rise from MJD 55804, the source occupies a Low/Hard state. The source is definitely in a hard-intermediate state (HIMS) on MJD 55818 as indicated by the hardness ratio, though where it enters the HIMS cannot be exactly determined. It continues in the hard-intermediate state till MJD 55840 after which the hardness ratio remains $\geq$ 1, and the source thus decays to a Low/Hard state.

\section{Discussions and conclusions}
\label{sec:discu}
It is evident from the spectral and timing analysis that the source MAXI J1836$-$194 exhibits spectral state transition. During the rising phase, the source probably occupies a Hard state on MJD 55804, although the vertical portion of the HID diagram (Figure \ref{HID-rms}) is not observed. From the same figure it is clear that the source travels along an almost horizontal track towards lower hardness ratios. Around MJD 55818 hardness ratio is $<$ 0.6 and clearly the source is in a hard-intermediate state. The return to hard state after MJD 55840 is more like a typical decay of HID to quiescent state.  

The above state transition is also supported by the fact that there is a disk contribution to the spectral fit also. Figure \ref{flux-frac} clearly shows that the contribution of disk component attains a maximum of 16.5\%. The thermal flux contribution never reaches anywhere close to the 80\%  which is typically observed in Black hole binaries while in a fully (canonical) soft state \citep{RM06,TB2010,Dunn2011}. Although the hard power law component dominates the total flux contribution throughout the outburst, clearly the presence of the soft component is necessary in the spectral fit. This is further strengthened by the contribution of the Fe line flux albeit small.  

It may also be noted that the contribution of soft component flux is much higher than observed in other sources like SWIFT J1753.5$-$0127 \citep{MCRS07,Zhang2007,Chiang2010} and XTE J1118$+$480, GRO J0422$+$22 \citep{Brock2004},  which remain in the hard state throughout their outbursts and are termed as `failed' outbursts.. Similar spectral analysis have been conducted  by \citealt{Ferrigno2012} for MAXI J1836$-$194, and have termed this as a failed outburst. In our analysis we stress the feature that the source did reach the Hard intermedite state and this is supported by the concurrent temporal features also.
 
The trend of increase in break frequency is another parameter which is indicative of the source entering the Hard intemediate state as seen in Figure \ref{break-flux},  and is also supported by the total rms values of the PDS as shown in Figure \ref{HID-rms}.  However, it is also observed that the break frequency instead of increasing further as would be expected of a source entering soft state, it starts to decrease after MJD 55820. The QPO frequency and its Q- factor also are highest on the same day.  

The temporal evolution of the source shows that the break frequency evolves with time. Panel C of Figure \ref{lc} shows that the hard photons reduce drastically on MJD 55820 when the break frequency has reached its maximum. This suggests the increase in soft flux, which has been observed simultaneously in Figure \ref{break-flux} as mentioned in \S \ref{sec:res}. Previous studies of GX 339$-$4 by \citealt{Plant2013} have shown that the break frequency increases while in the intermediate states, which is similar to what we observe in MAXI J1836$-$194. Until now different theories have been developed to understand the nature of break frequency evolution \citep{ST95,Titarchuk2007,ID2011}. Almost all the theories connect the origin of the break frequency with fluctuations in the inner boundary of a Keplerian disk. \citealt{Plant2013} compared the evolution of inner disk radius and break frequency as suggested by \citealt{Gilf99,Chura2001}. But we are unable to do the same for this outburst, since the source does not enter a predominant soft state. Evolution of the break frequency suggests that some oscillation occurs in the region where the Keplerian flow gets truncated. Figure \ref{break-flux} shows that the break frequency is maximum when the disk flux is around its peak.  
  
Quasi periodic oscillations (QPOs) are observed only for very few observations during the outburst as seen in Figure \ref{qpo_evo}. During this period, the inner disk temperature T$_{in}$ varies from 0.62 keV to $\approx$ 0.8 keV and the photon index is $\sim$ 1.9. This is when the source exists in the hard intermediate state. The QPOs are seen when the disk flux has reached a maximum and the spectral index has also increased. The variations in different temporal parameters over days have been summarized in the table \ref{t1}. QPOs are observed during MJD 55819 to 55822. The observation on MJD 55823 does not show the presence of QPOs. The next two observations on MJD 55824 and MJD 55825 show significant QPOs in their PDS with decreased peak frequency. It is interesting to note that the QPO frequencies observed in low/hard transients by \citealt{Brock2004,MCRS07} have been in the range of mHz to 1 Hz, while for MAXI J1836$-$194 we find that the QPO frequency is $\sim$ 3 to 5 Hz, which is higher than those observed for other low/hard transients. For sources which exhibit canonical outburst, QPO frequencies observed are up to 10 Hz (e.g. XTE J1859$+$226; \citealt{Casella2004}), or occassionally even more extending to 20 Hz(e.g. XTE J1748$-$288; \citealt{Rev99}).

This source therefore appears to be the second source which entered the HIMS but does not reach the soft state. Similar features have been observed in H 1743$-$322 during its October 2008 outburst \citep{Capitanio2009, Motta2010}. This source therefore appears to be the second source which entered the HIMS but does not reach the soft state. Both MAXI J1836$-$194 and H 1743$-$322 enter a hard intermediate state and exhibit low frequency QPOs in the few Hz range. These are the two sources which have exhibited HIMS after the hard state, as compared to other sources with failed outbursts which remain in hard state only (SWIFT J1753.5$-$0127; \citealt{MCRS07,Zhang2007,Chiang2010}, XTE J1118$+$480, GRO J0422$+$22; \citealt{Brock2004}). MAXI J1836$-$194 was softest on MJD 55820 and H 1743$-$322 was softest on MJD 54764. One other curious feature of the two sources is that the QPOs are not observed on one day (MJD 55823 for MAXI J1836$-$194 and MJD 54764 for H 1743$-$322) and after that the soft component decreases for both the sources. For MAXI J1836$-$194, a radio flare was observed at 2011 September 20 09:50:24.0 UT on MJD 55824.42 by RATAN-600\citep{Trushkin}, while for H 1743$-$322 there is no report of a peak radio flare close to MJD 54764 although there has been detection of a Radio counterpart on 2008 October 7 \citep{CorbelH08}. Observations and results suggests that the observation of QPO not being observed in the PDS on MJD 55823 is probably correlated with the subsequent detection of a Radio flare of 50 mJy on MJD 55824.42. There have been reports of QPO being not observed in the PDS of Black hole binaries whenever a flare occurs (See \citealt{MJ2012, RNS2013, NRS2013}). Detailed analysis on this will be performed to understand the phenomenon of disk-jet coupling for the source MAXI J1836$-$194, in the context of the Two Component Advective Flow (TCAF) model \citep{ST95} in the presence of magnetic field \citep{Nandi2001} and presented elsewhere (\citealt{RNS}).
 
We conclude from the analysis of RXTE observations of MAXI J1836$-$194 that 

\begin{itemize}
\item The source exhibits state transition of hard state $\rightarrow$ hard intermediate state $\rightarrow$ hard state
\item The source never occupies a soft state during the outburst but does reach HIMS, and therefore belongs to the category of few sources which enter HIMS but do not reach the soft state. 
\item QPOs are observed for a short duration when the source exists in the hard intermediate state
\item The evolution of break frequency is also observed which is linked with the disk characteristics 
\end{itemize}

\section*{Acknowledgments}

We are very thankful to Dr. Anuj Nandi of ISRO Satellite Centre, for relevant discussions based on the TCAF model and various suggestions. We would like to thank Prof. B. R. S. Babu of University of Calicut for his guidance. We acknowledge the research fellowship and research facilities provided by ISRO Satellite Centre, INDIA.

This research has made use of the data obtained through High Energy Astrophysics Science Archive Research Center online service, provided by NASA/Goddard Space Flight Center and the MAXI data provided by RIKEN, JAXA and the MAXI team. We have also used the General High-energy Aperiodic Timing Software (GHATS) package developed by T.M. Belloni at INAF - Osservatorio Astronomico di Brera.

\clearpage

\end{document}